# Spinel Cu-Mn-Cr Oxide Nanoparticle-Pigmented Solar Selective Coatings Maintaining >94% Efficiency at 750ºC

*Can Xu, Xiaoxin Wang, and Jifeng Liu\**

Thayer School of Engineering, Dartmouth College, 14 Engineering Drive, Hanover, New Hampshire 03755, USA

*Corresponding Author: Jifeng.Liu@dartmouth.edu

**ABSTRACT**

High-temperature concentrating solar power (CSP) system is capable of harvesting and storing solar energy as heat towards cost-effective dispatchable solar electricity. Solar selective coating is a critical component to boost its efficiency by maximizing solar absorptance and minimizing thermal emittance losses. However, maintaining a high solar-thermal conversion efficiency >90% for long-term operation at ≥750ºC remains a significant challenge. Herein, we report spray-coated spinel Cu-Mn-Cr oxide nanoparticle-pigmented solar selective coatings on Inconel tube sections maintaining ≥94% efficiency at 750ºC and ≥92.5% at 800ºC under 1000x solar concentration after 60 simulated day-night thermal cycles in air, each cycle comprising 12h at 750ºC/800ºC and 12h cooling to 25ºC. The solar spectral selectivity is intrinsic to the band-to-band and d-d transitions of non-stoichiometric spinel Cu-Mn-Cr oxide nanoparticles by balancing the lattice site inversion of $Cu^{2+}$ and $Mn^{3+}$ on tetrahedral vs. octahedral sites. This feature offers a large fabrication tolerance in nanoparticle volume fraction and coating thickness, facilitating low-cost and scalable spray-coated high-efficiency solar selective absorbers for high-temperature CSP systems.

**Key Words:** Concentrating solar power, Solar selective absorber, Spinel oxide nanoparticle; Ionic site inversion; d-d transition,

**TOC GRAPHICS**

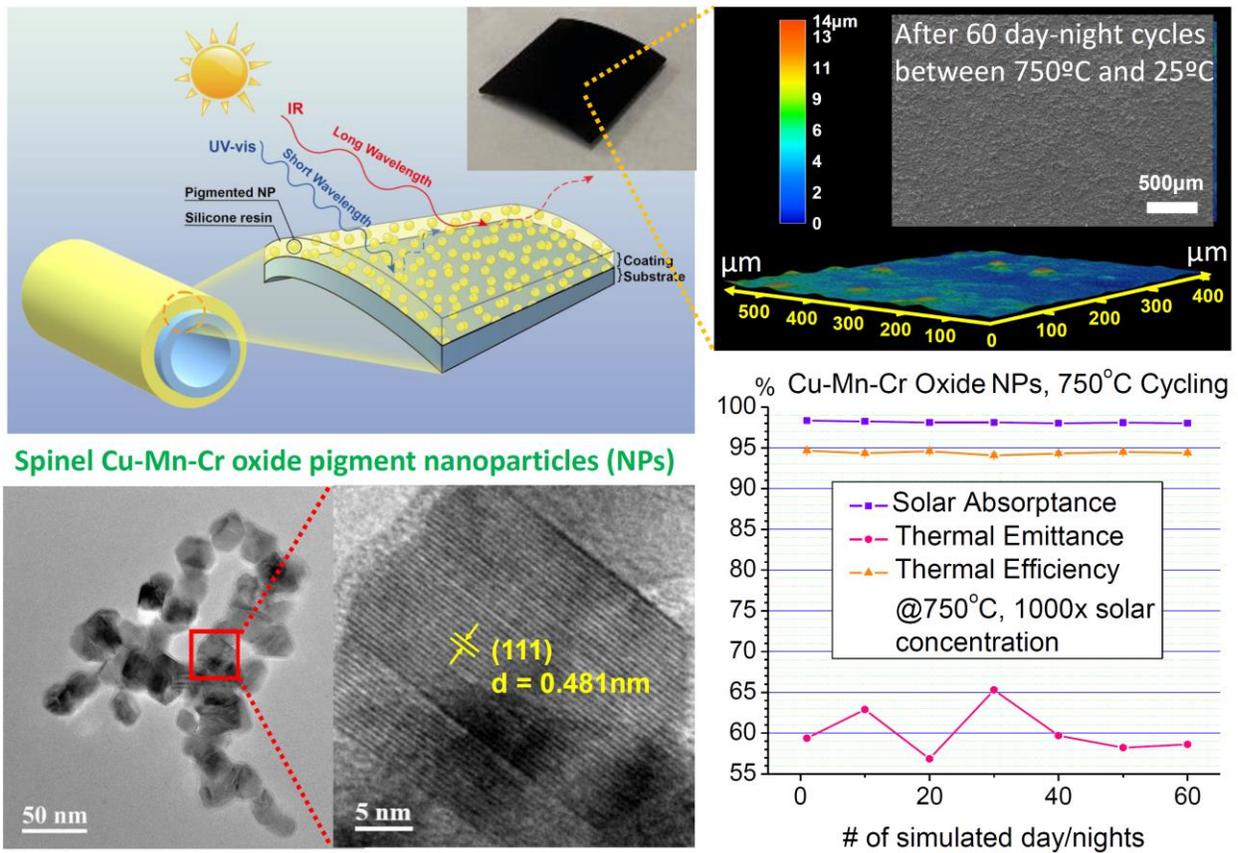

## 1. Introduction

Recent years have seen a rapid growth in solar energy, currently supplying 2.8% of electricity in the U.S. as the third largest renewable source after wind and hydropower. [1] Concentrating solar power (CSP) system utilizes reflective mirrors ("collectors") to concentrate solar irradiation and heat up working fluids (e.g. molten salts) to high temperatures. [2] A great advantage compared to photovoltaic (PV) system is that CSP is capable of storing the solar-thermal energy for >10 hours so as to meet the peak hours of electricity consumption towards dispatchable solar electricity. [3] Solar selective coating, a critical component to boost the solar-to-thermal energy conversion efficiency ($\eta_{therm}$) by maximizing solar spectral absorption and minimizing infrared (IR) thermal emittance losses, can reduce the levelized cost of energy (LCOE) of CSP by >12% [4] for $\eta_{therm}$>90%. Further increasing $\eta_{therm}$ to 95% is projected to achieve >17% LCOE cost reduction, which strongly supports the goal of achieving $0.05/kWh solar electricity by 2030. [3] Based on Carnot Theorem, the operation temperature of Generation 3 CSP system is being increased to 750ºC with a solar concentration of ~1000x to achieve >50% overall power cycle efficiency in solar electricity production. [5] Therefore, it is highly desirable to develop cost-effective and highly scalable solar selective coatings that can maintain $\eta_{therm}$ ~95% at 750ºC in air.

However, currently commercial solar coating products are unable to maintain $\eta_{therm}$ >90% at operating temperatures >700ºC in air. [6] The benchmark Pyromark 2500 coating suffers from a high thermal emittance of $\varepsilon_{therm}$~87% and pigment particle phase instability at 750°C, [7] limiting its $\eta_{therm}$ to ~88% at 1000x solar concentration after operating at 750°C for 300 h. [8,9] Various transition metal oxide solar absorbers have been investigated, and some of them demonstrate excellent thermal stability at 750°C, [10,11] yet the lack of spectral selectivity limits their maximal $\eta_{therm}$ to ~91% at >700°C. While dielectric selective absorber coatings based on nitrides and oxides could sustain high temperature in air and achieve some degree of solar selectivity, [12,13] the principles of their optical design requires relatively expensive vacuum deposition for stringent thickness control.

Our previous optical design [14] based on Lorentz-Mie scattering theory and Four-Flux model has found it feasible to achieve good solar selectivity in nanoparticle (NP)-pigmented silicone coatings for NPs with diameters <40 nm and steep optical transition near the optimal cut-off wavelength ($\lambda_{cut}$) between the solar spectrum and the thermal radiation spectrum. For Generation 3 CSP systems, $\lambda_{cut}$=2475 nm for 1000x solar concentration at 750ºC. Spinel ($AB_2O_4$) NPs with intrinsic high-temperature thermal stability is a promising candidate since their manifold compositions and cationic valences enable a high degree of freedom to tailor the optical properties. Previously, we have demonstrated $MnFe_2O_4$ NP-pigmented solar selective coatings maintaining $\eta_{therm}$ ~89% under 1000x solar concentration after serving at 750ºC in air for 700h. [15] To further enhance the performance, in this Paper we demonstrate spinel Cu-Mn-Cr oxide NP-pigmented silicone solar selective coatings on Inconel tube sections with a higher solar absorbance $\alpha_{solar}$=98.2% and a notably reduced thermal emittance $\varepsilon_{therm}$=59.4% compared to benchmark Pyromark 2500 ($\alpha_{solar}$=96.0%; $\varepsilon_{therm}$=89.5%). To the best of our knowledge, this performance leads to a record-high $\eta_{therm}$=94.5±0.2% for 1000x solar concentration at 750ºC in air. These coatings also maintain $\eta_{therm}$≥94% at 750ºC and $\eta_{therm}$ ≥92.5% at 800ºC after 60 simulated day (750ºC/800 ºC 12h) and night (25 ºC 12h) cycles in air without degradation in surface morphology or phase stability. The solar spectral selectivity is intrinsic to the band-to-band and d-d transitions of non-stoichiometric spinel Cu-Mn-Cr oxide NPs. This feature offers a large fabrication tolerance in nanoparticle volume fraction and coating thickness, greatly facilitating high-efficiency solar selective absorbers layers via low-cost and highly scalable spray coating for high-temperature CSP systems.

## 2. Results and Discussions

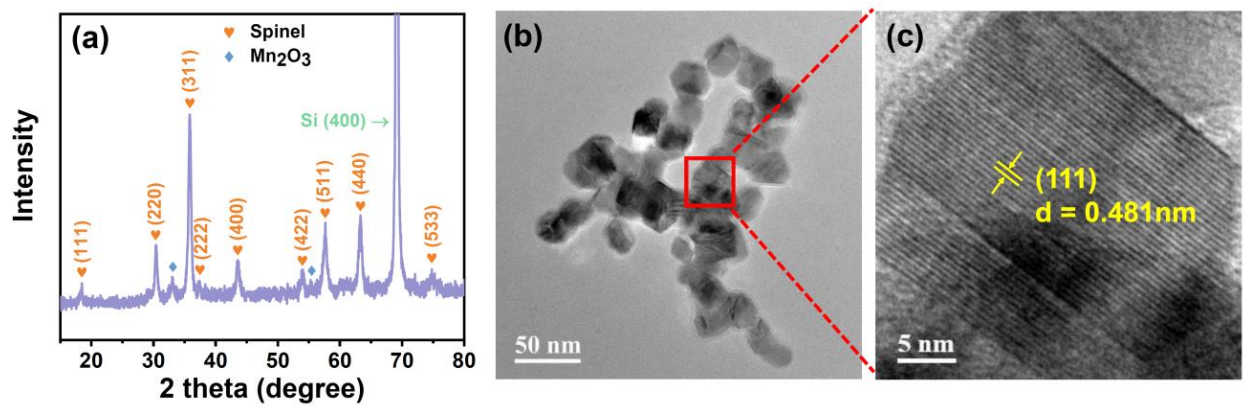

**Figure 1**. (a) XRD pattern of the synthesized spinel Cu-Mn-Cr oxide NPs. A small amount of $Mn_2O_3$ is also identified. (b) shows a TEM image of the NPs, and (c) zooms into one of the NPs in the red box shown in (b).

**Structural and Compositional Analyses of Spinel Cu-Mn-Cr Oxide NPs**. We utilized co-precipitation method to synthesize spinel oxide NPs with Cu:Mn:Cr=1:3:1 from nitrate salt precursors (see Supporting Information Section 1). Such a nonstoichiometric ratio is selected to optimize the solar selective absorption by engineering the valences of cations and cationic site distribution to our advantage, as will be discussed later. These NPs are further annealed at 550°C to enhance the crystallinity. Energy dispersive X-ray spectroscopy (EDS) analysis shows Cu:Mn:Cr =1.00:3.33:1.17 in the synthesized NPs, close to the targeted ratio. **Figure 1a** shows the X-ray diffraction (XRD) pattern of the synthesized NPs loaded on Si(100) substrate. Most of the peaks are attributed to spinel structure with a lattice constant of $a$=8.307 Å, between those of cubic $CuMn_2O_4$ ($a$=8.331 Å) and $CuCr_2O_4$ ($a$=8.270 Å) [16] as expected for spinel Cu-Mn-Cr oxide. A couple of small peaks from $Mn_2O_3$ are also observed, consistent with the phase diagram reported for Cu-Mn spinel oxides with Cu:Mn <1. [17] Based on XRD relative intensity ratio (RIR) analysis, the molar ratio of Cu-Mn-Cr spinel oxide to $Mn_2O_3$ is 1:(0.142□0.018). **Figures 1b** and **1c** further show transmission electron microscopy (TEM) images of the NPs. The average NP diameter is 31±3.6 nm (see the NP size histogram in Supporting Information Figure S1). The interplanar spacing of {111} planes is 4.81 Å in the high-resolution TEM image in Figure 1c, fully consistent with the XRD results.

Furthermore, X-ray photoelectron spectroscopy (XPS) analyses provide the valences and the corresponding percentages of the cations, as summarized in Table 1 (see Supporting Information Section 3 for data analyses). Notably, the $Cu^+:Cu^{2+}$ ratio is as high as 4:1. It has been reported that $Cu^+$ on tetrahedral A sites in spinel oxides can be stabilized by $Cu^{2+}$ and $Mn^{4+}$ on octahedral B sites [17]. According to ligand field theory (LFT) and the octahedral site preference energy (OSPE) [18,19], $Cu^+(3d^{10})$ and $Mn^{2+}(3d^5)$ prefer tetrahedral sites while $Cu^{2+}$, $Mn^{3+}$ and $Cr^{3+}/Mn^{4+}$ are sequentially more energetically favored to take the octahedral sites. Therefore, considering the cation valance distribution, OSPE, and overall charge balance, the detailed formula can be written as $(Cu^+_{0.48}Mn^{2+}_{0.52})_{Td,A}(Cu^{2+}_{0.12}Mn^{2+}_{0.07}Mn^{3+}_{0.65}Mn^{4+}_{0.46}Cr^{3+}_{0.70})_{oh,B}O_{3.89}$ assuming OSPE fully applies. Here, the subscripts "Td,A" and "Oh,B" stand for tetrahedral A site and octahedral B site, respectively. Characteristic vibration modes corresponding to octahedral and tetrahedral sites in spinel structures are also observed in Fourier transform IR spectroscopy (FTIR) and detailed in the Supporting Information (Figure S3). On the other hand, even though OSPE is highly effective in predicting cation lattice sites, we note that ~30% site inversion of $Mn^{3+}$ from octahedral to tetrahedral sites has been reported in closely related $CuMn_2O_4$ spinel structures in order to dilute the Jahn-Teller effect. [20] Such $Mn^{3+}$ site inversion can lead to strong and broad absorption bands in the near infrared (NIR) regimes at 1000-2000 nm wavelength. [21,22] Furthermore, the deficiency of oxygen compared to stoichiometric $AB_2O_4$ indicates oxygen vacancies, likely on the surface of the NPs as has been reported in other spinel NP systems. [23] Oxygen vacancies are known to interact with transition metal cations and further enhanced the NIR absorption. [24] These influences on optical properties will be discussed next.

**Table 1**. Cu:Mn:Cr atomic ratios and corresponding cationic valences/percentages in the synthesized NPs comprising both spinel Cu-Mn-Cr oxide and $Mn_2O_3$ at a ratio of 1:0.142

| Atomic Species | Atomic Ratio | Ionic Valence/Molar Percentage |
|---|---|---|
| Cu | 1.00 | $Cu^+$: 80% |
|  |  | $Cu^{2+}$: 20% |
| Mn | 3.33 | $Mn^{2+}$: 30% |
|  |  | $Mn^{3+}$: 47% |
|  |  | $Mn^{4+}$: 23% |
| Cr | 1.17 | $Cr^{3+}$: 100% |

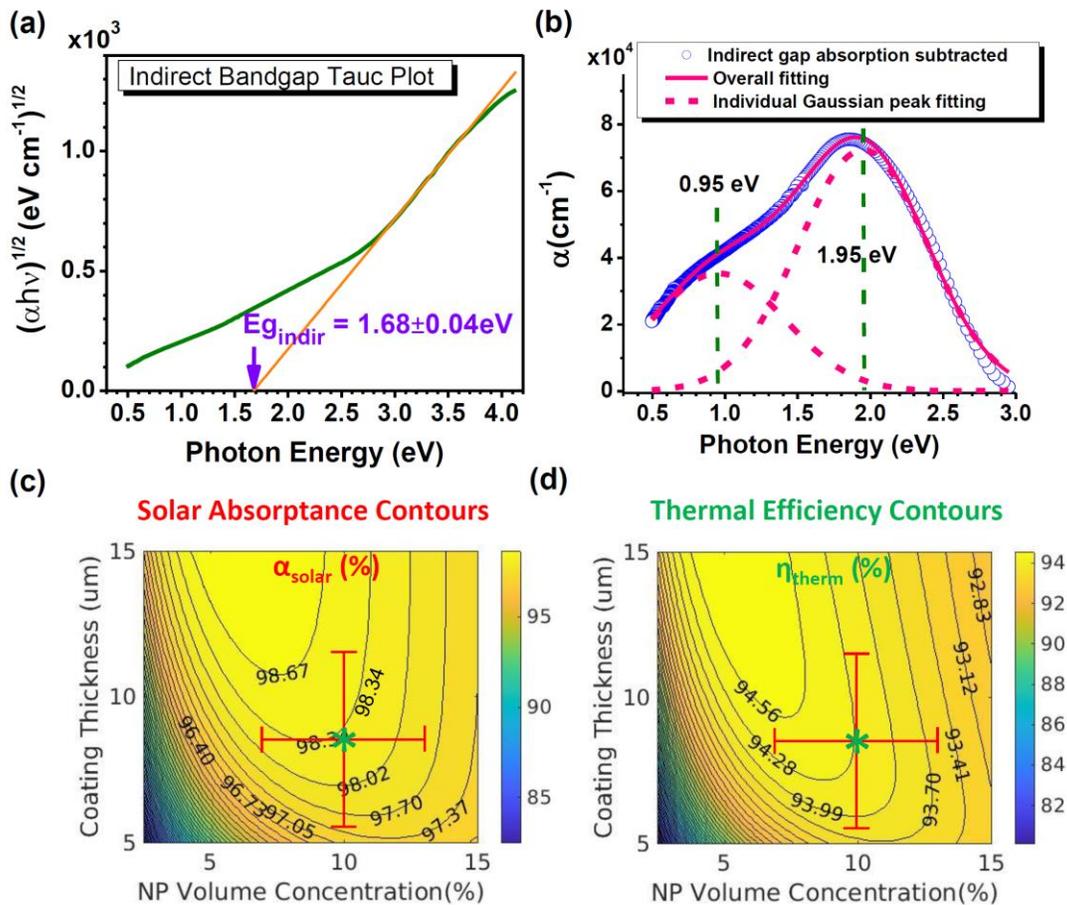

**Figure 2**. (a) Indirect bandgap Tauc plot of the spinel Cu-Mn-Cr oxide NPs (b) Absorption spectrum beyond the indirect bandgap and the corresponding Gaussian peaks fitting. The two

absorption bands peaked at 0.95 eV and 1.95 eV are mainly contributed by $Cu^{2+}$ and $Mn^{3+}$ d-d transitions on tetrahedral and octahedral sites, respectively. (c) and (d) show theoretical modeling of solar absorptance and thermal efficiency contour maps as a function of NP volume fraction and coating thickness when dispersed in silicone matrix.

**Optical Properties**. Critical to the optical performance is the absorption spectrum of Cu-Mn-Cr oxide NPs, as shown in the Tauc plot in **Figure 2a**. It reveals an indirect gap of 1.68□0.04 eV. Remarkably, the absorption beyond the indirect gap is extended all the way to 0.5 eV with absorption coefficients >$10^4$ cm$^{-1}$ to cover the entire solar spectrum. To single out the absorption spectrum beyond the indirect bandgap, we subtract the indirect bandgap absorption (as derived from the Tauc plot) from Figure 2a and show the result in **Figure 2b**. Two broad Gaussian absorption bands peaked at 0.95 and 1.95 eV are clearly identified, typical of d-d transitions between the split d levels of transition metal ions induced by tetrahedral or octahedral ligand (crystal) field. [21] The energy ratio of these two peaks is 0.48, very close to the expected ratio of 4/9 for tetrahedral site vs. octahedral site d-d transition energies in spinel structures. [25] Therefore, the 0.95 eV and 1.95 eV absorption bands are attributed to tetrahedral and octahedral site d-d transitions, respectively. Note that in most of these d-d transitions, the excited electrons still remain localized to the transition metal ion instead of becoming free electrons in the conduction band, therefore the transition energy is lower than the bandgap. The 0.95 eV peak is between the tetrahedral site $Cu^{2+}$ absorption band peaked at ~0.8 eV [25–27] and that of $Mn^{3+}$ peaked at ~1.2 eV, [21,22] suggesting both types of ions on tetrahedral sites have contributed to this NIR d-d absorption band. Typically, the oscillation strength of tetrahedral d-d transition is stronger than their octahedral counterparts due to broken inversion symmetry that enables various spin-forbidden transitions.[21,25] In our case, on the other hand, the octahedral d-d absorption at 1.95 eV is ~2x stronger than tetrahedral absorption at 0.95 eV. This result indicates that a relatively small fraction of $Cu^{2+}$ and $Mn^{3+}$ cations occupy tetrahedral sites compared to octahedral sites, consistent with the prediction of OSPE discussed earlier and similar to the case of $CuMn_2O_4$ in terms of a small fraction of site inversion.[20] Such a distribution is beneficial to solar selectivity, where a decrease in absorption beyond the optimal cut-off wavelength of 2475 nm ($hv$<0.5 eV) is needed, as mentioned earlier. Since $Cr^{3+}$ and $Mn^{4+}$ have the strongest tendency to compete for octahedral sites based on OSPE, tuning their percentages may further optimize $Cu^{2+}$ and $Mn^{3+}$ site inversion for better solar selectivity. In addition, oxygen vacancies, as identified in our previous analyses, further lower the ligand symmetry to enhance the oscillation strength of d-d optical absorption especially in the NIR solar spectral regime at 0.5-1 eV, as has been reported in spinel $ZnFe_2O_4$ system.[28] Therefore, these two factors work synergistically to enhance the overall solar selectivity.

**Optical Modeling of Spinel Cu-Mn-Cr Oxide NP-Pigmented Solar Selective Coatings**. Based on the absorption spectra of the spinel Cu-Mn-Cr oxide NPs, we further modelled the spectrally integrated solar absorptance $\alpha_{Solar}$ and the thermal efficiency $\eta_{therm}$ for 1000x solar concentration at 750ºC using Lorentz-Mie scattering theory and four-flux radiative model, as detailed in Refs. 14 and 15. **Figure 2c** and **d** show $\alpha_{Solar}$ and $\eta_{therm}$ contour maps, respectively, as a function of NP volume fraction and coating thickness when dispersed in silicone matrix for the solar absorber coating. The intrinsic solar spectral selectivity of the NPs and their nanoscale diameters (d=31±3.6 nm) lead to a large tolerance in coating thickness and NP volume fraction, such that $\alpha_{Solar}$>97.4% and $\eta_{therm}$>93.6% can be achieved even if the NP volume fraction varies between 7-13 vol.% and the coating thickness varies between 6-15 μm. Such a high fabrication tolerance greatly facilitates low-cost and highly scalable spray-coated solar selective absorbers. The green stars and red error bars in Figures 2c and d reflect the experimentally measured coating thickness and NP volume fraction variations in our spray-coated Inconel tube section samples. From the modeling, we expect $\alpha_{Solar} \approx 98\%$ and $\eta_{therm} \approx 94\%$. As will be detailed next, these theoretically modelled values indeed agree very well with the experimental results.

**Characterization and Optical Performance of the Spray-Coated Solar Selective Coating.**
**Figure 3a** shows a photograph of an Inconel 625 tube section (outer diameter=76 mm) coated with the spinel Cu-Mn-Cr oxide NP-pigmented silicone solar selective absorber and annealed in air at 750ºC for 24h. Detailed coating procedure is discussed in Section 1 of the Supporting Information. **Figure 3b** shows a digital optical microscopy image of the surface profile, indicating a surface undulation of ~6 μm with sporadic humps reaching >10 μm in height. A focus ion beam (FIB) cross-sectional cutting is made on a relatively flat and thin region, revealing an average thickness of 5.5 μm as shown in **Figure 3c**. Using this flat region as a calibration and the surface profile in Figure 3b, we determined that overall, the coating thickness is 8.5±3 μm. The volume fraction of the spinel Cu-Mn-Cr oxide NPs is estimated to be 10±3 vol.%, as detailed in Section 5 of the Supporting Information. These parameters allow us to model the expected performance of the coating, as shown in Figures 2c and 2d. The XRD data in **Figure 3d** shows that the spinel structure is well maintained and the crystallinity gets better (narrower diffraction peaks) after 750ºC annealing for 24h in air.

Remarkably, **Figure 3e** and **f** demonstrate that the new spinel Cu-Mn-Cr oxide NP-pigmented solar selective coating improves the solar absorptance $\alpha_{Solar}$ from 96.0% to 98.2% compared to benchmark Pyromark 2500, while simultaneously the thermal emittance $\varepsilon_{therm}$ is drastically reduced from 89.5% to 59.4%. Correspondingly, $\eta_{therm}$ is notably increased from 90.4±0.3% to 94.5±0.2%. As will be further discussed in Table 2, to the best of our knowledge, so far this is the highest optical-to-thermal conversion efficiency for air-stable solar selective coatings operating at 750ºC.

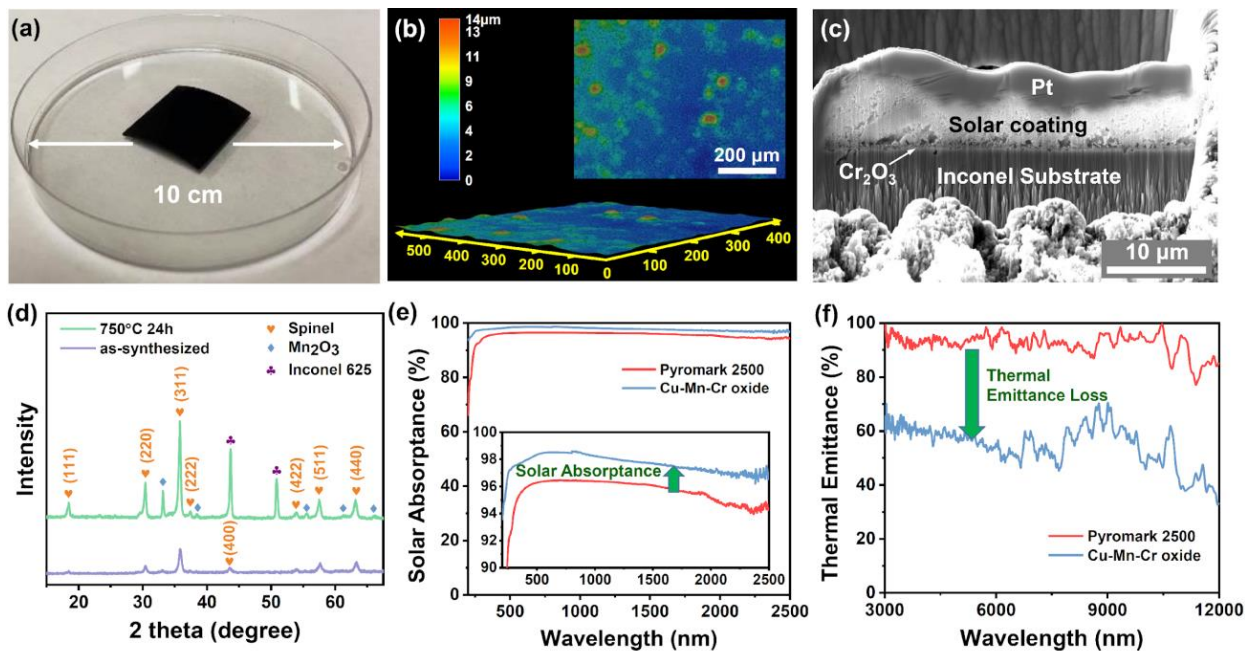

**Figure 3**. (a) A photograph of an Inconel 625 tube section (outer diameter=76 mm) coated with spinel Cu-Mn-Cr oxide NP-pigmented silicone solar selective absorber layer. (b) An optical topography map of the solar selective coating, including a 3D view and a top view in the inset. (c) FIB cross-section of the coating in a relatively flat and thin region in (b) for thickness calibration. (d) XRD pattern of the coated sample after annealing for 24 h at 750ºC in air compared to that of the as-synthesized particles. (e) and (f) show solar absorptance and thermal emittance spectra of the Cu-Mn-Cr oxide NP-pigmented silicone solar selective coating compared to benchmark Pyromark 2500.

**Endurance Testing.** We further performed extensive thermal cycling for endurance testing of these high efficiency solar selective coatings. Each simulated day/night thermal cycle includes 12h annealing at 750ºC or 800ºC and 12h cooling to 25ºC. Here thermal cycles at 800ºC (i.e., 50ºC higher than the working temperature) are conducted to further confirm thermal stability and to investigate possible degradation mechanisms at an accelerated rate. Solar coatings on Inconel 625 tube sections (with outer diameter=76 mm as shown in Figure 3) are annealed for up to 60 simulated day/night cycles in air. Comparing the surface morphology from scanning electron microscopy (SEM) images shown in **Figures 4a-c**, we find no deterioration in coating integrity or appreciable changes in morphology after 60 cycles between 750ºC/800ºC and 25ºC. The micropores on the surface of the samples help to accommodate volume changes upon thermal cycling, thereby stabilizing the coating against thermal stress. Such surface texture also helps to reduce surface reflectance and enhance solar absorption, similar to the case of PV cells. XRD analyses further show that the spinel Cu-Mn-Cr oxide NPs are thermodynamically stable upon thermal cycling (see Section 6 of the Supporting Information). **Figures 4d-f** show the evolution of solar absorptance spectra, thermal emittance spectra, and spectrally integrated solar absorptance/thermal emittance/thermal efficiency vs. the number of thermal cycles between 750ºC and 25ºC. The solar absorptance decreases only very slightly after 60 thermal cycles between 750ºC and 25ºC, while the thermal emittance fluctuates around 60% during the cycling, maintaining a record-high thermal efficiency $\eta_{therm}$>94% during the entire thermal cycles. Similar data shown in **Figures 4g-i** indicate a more noticeable decrease in solar absorptance when the high temperature cycles are increased to 800ºC. Even so, a high $\eta_{therm}$=92.8±0.3% is still maintained after 60 thermal cycles between 800ºC and 25ºC. The increasingly wavy solar absorption spectra upon 800ºC/25ºC thermal cycling with reduced solar absorptance closely resemble the behavior of $CuCr_2O_4$ NP-pigmented coatings (intentionally synthesized for comparison; see Section 7 of Supporting Information) as well as previous literature on $CuCr_2O_4$ due to $Cr^{3+}$ d-d absorption bands.[29] In fact, these peaks and valleys blueshift towards those of $CuCr_2O_4$ after more cycles. This result suggests Cr diffusion and substitution into the Cu-Mn-Cr spinel oxide NPs from the Inconel substrate as the key mechanism for the slight solar absorptance degradation upon 800ºC/25ºC thermal cycling. This is indeed confirmed by detailed cross-sectional EDS mapping before and after the thermal cycling, as detailed in Section 7 of the Supporting Information. Therefore, limiting Cr diffusion into the coating could further improve the endurance of the solar selective coating. A possible approach is to pre-oxidize the Inconel substrate to form a $Cr_2O_3$ layer first before spray-coating, which has proved to be an effective approach to address the Cr diffusion issue in our previous work. [15]

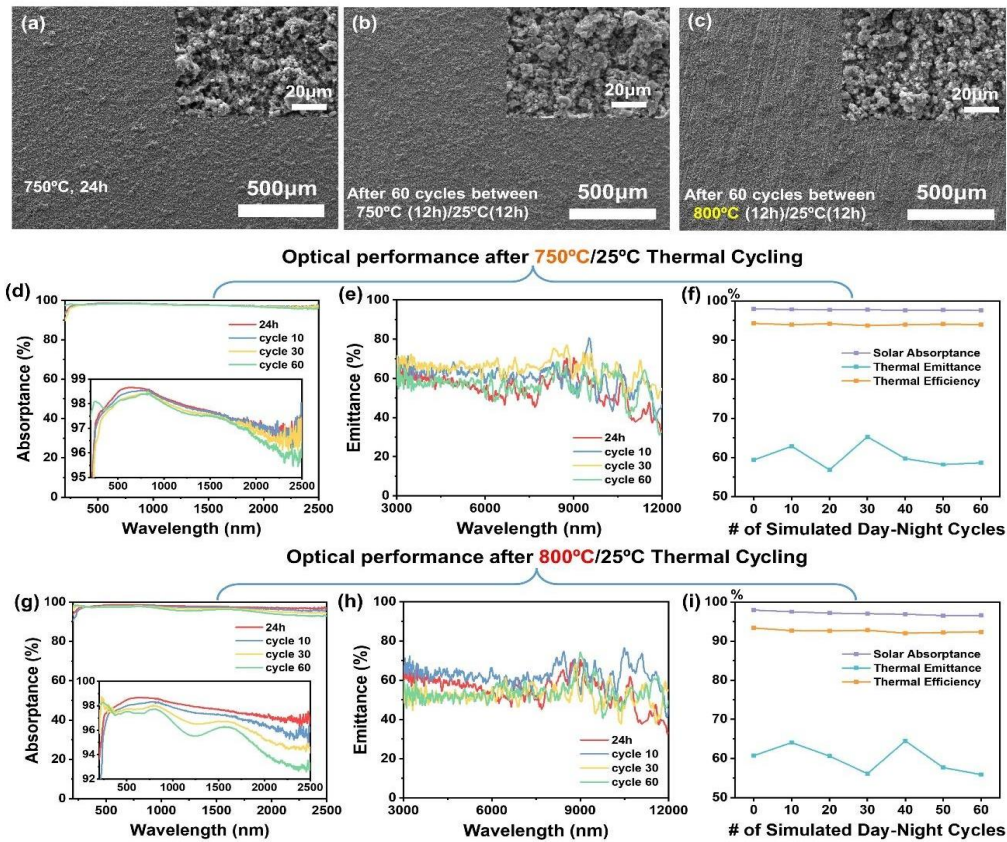

**Figure 4**. SEM surface morphology of the coatings on Inconel 625 tube sections (76 mm outer diameter) after (a) 750ºC 24h annealing; (b) 750ºC 24h annealing plus 60 simulated day-night

cycles between 750 ºC and 25 ºC; (c) 750ºC 24h annealing plus 60 simulated day-night cycles between 800 ºC and 25 ºC. (d)-(f) show the evolution of solar absorptance spectra, thermal emittance spectra, and spectrally integrated solar absorptance/thermal emittance/thermal efficiency vs. the number of thermal cycles between 750ºC and 25ºC. (g)-(i) show similar data for thermal cycles between 800ºC and 25ºC

Table 2 compares the efficiency and endurance of our coating with some recent work. Most of the previous solar coatings lack spectral selectivity with a thermal emittance ~90%, limiting their thermal efficiency to $\eta_{therm}$~90.5%. We have notably improved $\eta_{therm}$ to >94% by engineering and balancing the intrinsic NIR vs. visible d-d absorption bands of $Cu^{2+}$ and $Mn^{3+}$ on tetrahedral vs. octahedral sites of spinel structure. The thermal emittance is drastically reduced to ~60% while a high solar absorptance of ~98% is maintained. To the best of our knowledge, this is the highest efficiency demonstrated and maintained so far for 750ºC endurance testing in air. Optimizing the extent of lattice site inversion of $Cu^{2+}$ and $Mn^{3+}$ on tetrahedral vs. octahedral sites by fine-tuning $Cr^{3+}$ and $Mn^{4+}$ percentages may further improve the spectral selectivity towards $\eta_{therm}$>95%. The intrinsic solar spectral selectivity of the spinel Cu-Mn-Cr oxide NPs also enables an excellent fabrication margin for cost-effective, highly scalable spray coating, as preliminary demonstrated on a 48-inch-long tube shown in Figure S8 of the Supporting Information.

**Table 2** High-temperature solar selective coatings with reported endurance test

| Material System | Substrate | Fabrication Method | $\eta_{start}$ | $\eta_{end}$ | T (°C) | Endurance in Air (h/°C) | Refs. |
|---|---|---|---|---|---|---|---|
| $Cu_{0.15}Co_{2.84}O_4$-SPB-$SiO_2$ | Inconel 625 | Spray Coating | 0.904 | 0.903 | 750 | 1000/750 | Ref.30 |
| $Cu_{1.5}Mn_{1.5}O_4$-SPB-$SiO_2$ | Inconel 625 | Spray Coating | 0.909 | 0.905 | 750 | 1000/750 | Ref.30 |
| Porous $Cu_{0.5}Cr_{1.1}Mn_{1.4}O_4$-$SiO_2$ | Haynes 230 | Spray Coating | 0.903 | 0.902 | 800 | 2000/800 | Ref.11 |
| $Cu_{0.86}Cr_{0.14}Mn_{1.5}Fe_{0.5}O_4$-$SiO_2$ | Inconel 617 | Spray Coating | ≤ 0.917 | ≤ 0.894 | 750 | 1300/800 | Ref.31 |
| TiN/AlCrSiO(two nano-multilayers)/AlCrSiO(amorphous) | SS | Cathode Arc Ion Plating | ≤ 0.908 | ≤ 0.867 | 750 | 200/700 | Ref.32 |
| Spinel Cu-Mn-Cr oxide NP-silicone | Inconel 625 | Spray Coating | **0.945** | **0.942** | 750 | 60 thermal cycles 750ºC/25 ºC | **This work-** |
| | | | **0.937** | **0.928** | 800 | 60 thermal cycles 800ºC/25 ºC | **This work-** |

$\eta_{start}$: efficiency as deposited; $\eta_{end}$: efficiency after annealing;

T: temperature at which thermal efficiency is evaluated.

Note that Refs. 31 and 32 only reported thermal emittance at 80ºC instead of high temperatures >700ºC. We therefore estimated the upper limit of the thermal efficiency at high temperatures in these cases using 80ºC thermal emittance values, considering thermal emittance typically increases at higher temperatures.

## 3. Conclusions

In conclusion, we demonstrate spray-coated spinel Cu-Mn-Cr oxide NP-pigmented solar selective coating that maintains $\eta_{therm}$ >94% upon 60 simulated day-night cycles between 750ºC and 25ºC in air. The spectral selectivity is intrinsic to the band-to-band and d-d transitions of these non-stoichiometric spinel NPs, where $Cu^{2+}$ and $Mn^{3+}$ on tetrahedral sites (through spinel site inversion) contribute to the NIR absorption band to cover the entire solar spectrum up to 2500 nm wavelength. This feature offers a large fabrication tolerance in NP volume fraction and coating thickness, greatly facilitating high-efficiency, high-temperature solar selective absorbers layers via low-cost and highly scalable spray coating for Generation 3 high-temperature CSP systems.

## ASSOCIATED CONTENT

**Supporting Information** includes experimental methods, nanoparticle size (diameter) histogram, XPS data analyses, FTIR data analyses, volume fraction determination, XRD analyses during thermal endurance tests, optical spectra and EDS mapping for interdiffusion investigation, and solar selective coating on a 48-inch-long tube.

## AUTHOR INFORMATION

**Corresponding Author**

Jifeng Liu − Thayer School of Engineering, Dartmouth College, 14 Engineering Drive, Hanover, New Hampshire 03755, United States; Email: jifeng.liu@dartmouth.edu

ORCID: 0000-0003-4379-2928

**Authors**

Can Xu − Thayer School of Engineering, Dartmouth College, 14 Engineering Drive, Hanover, New Hampshire 03755, United States; https://orcid.org/0000-0001-5306-5367;

Xiaoxin Wang − Thayer School of Engineering, Dartmouth College, 14 Engineering Drive, Hanover, New Hampshire 03755, United States;

**Notes**

The authors declare no competing financial interest.

## ACKNOWLEDGMENTS

This project is funded by U.S. Department of Energy, Solar Energy Technologies Office, under the award number DE-EE-0008530. We would like to thank Dr. Maxime J. Guinel at Dartmouth

College and Dr. Jules Gardener at Harvard University for their support with electron microscopy analyses, and Dr. Min Li at Yale University for support with XPS.## REFERENCES

(1)     EIA. What is U.S. electricity generation by energy source? https://www.eia.gov/tools/faqs/faq.php?id=427&t=3 (accessed 2022 -04 -13).

(2)     Jacobson, M. Z.; Delucchi, M. A. Providing All Global Energy with Wind, Water, and Solar Power, Part I: Technologies, Energy Resources, Quantities and Areas of Infrastructure, and Materials. *Energy Policy* **2011**, *39* (3), 1154–1169. https://doi.org/10.1016/j.enpol.2010.11.040.

(3)     Murphy, C.; Sun, Y.; Cole, W.; Maclaurin, G.; Turchi, C.; Mehos, M.; Murphy, C.; Sun, Y.; Cole, W.; Maclaurin, G.; Turchi, C.; Mehos, M. *The Potential Role of Concentrating Solar Power within the Context of DOE's 2030 Solar Cost Targets*; Golden, Colorado, 2019.

(4)     Boubault, A.; Ho, C. K.; Hall, A.; Lambert, T. N.; Ambrosini, A. Levelized Cost of Energy (LCOE) Metric to Characterize Solar Absorber Coatings for the CSP Industry. *Renew. Energy* **2016**, *85*, 472–483. https://doi.org/10.1016/j.renene.2015.06.059.

(5)     Mehos, M.; Turchi, C.; Vidal, J.; Wagner, M.; Ma, Z.; Ho, C.; Kolb, W.; Andraka, C.; Kruizenga, A. *Concentrating Solar Power Gen3 Demonstration Roadmap*; Golden, Colorado, 2017. https://doi.org/10.2172/1338899.
17

(6) Suman, S.; Khan, M. K.; Pathak, M. Performance Enhancement of Solar Collectors - A Review. *Renew. Sustain. Energy Rev.* **2015**, *49*, 192–210. https://doi.org/10.1016/j.rser.2015.04.087.

(7) Ho, C. K.; Ambrosini, A.; Bencomo, M.; Hall, A. C.; Lambert, T. N.; Mahoney, A. R. Characterization of Pyromark 2500 for High-Temperature Solar Receivers. In *the ASME 2012 6th International Conference on Energy Sustainability*; San Diego, CA, 2012.

(8) Ambrosini, A. High-Temperature Solar Selective Coating Development for Power Tower Receivers https://www.energy.gov/sites/prod/files/2016/08/f33/R 5 - Ambrosini_CSP-Program-Summit-SSC Presentation_to DOE.pdf.

(9) Ambrosini, A.; Lambert, T. N.; Boubault, A.; Hunt, A.; Davis, D. J.; Adams, D.; Hall, A. C. Thermal Stability of Oxide-Based Solar Selective Coatings for CSP Central Receivers. In *Proceedings of the ASME 2015 9th International Conference on Energy Sustainability*; San Diego, California, 2015; pp 1–10. https://doi.org/10.1115/ES2015-49706.

(10) Moon, J.; Kyoung Kim, T.; VanSaders, B.; Choi, C.; Liu, Z.; Jin, S.; Chen, R. Black Oxide Nanoparticles as Durable Solar Absorbing Material for High-Temperature Concentrating Solar Power System. *Sol. Energy Mater. Sol. Cells* **2015**, *134*, 417–424. https://doi.org/10.1016/j.solmat.2014.12.004.

(11) Rubin, E. B.; Chen, Y.; Chen, R. Optical Properties and Thermal Stability of Cu Spinel Oxide Nanoparticle Solar Absorber Coatings. *Sol. Energy Mater. Sol. Cells* **2019**, *195* (January), 81–88. https://doi.org/10.1016/j.solmat.2019.02.032.

(12) Cao, F.; McEnaney, K.; Chen, G.; Ren, Z. A Review of Cermet-Based Spectrally Selective Solar Absorbers. *Energy Environ. Sci.* **2014**, *7* (5), 1615–1627. https://doi.org/10.1039/c3ee43825b.

(13) Ambrosini, A. High-Temperature Solar Selective Coating Development for Power Tower Receivers http://energy.gov/sites/prod/files/2014/01/f7/csp_review_meeting_042413_ambrosini.pdf.

(14) Wang, X.; Yu, X.; Fu, S.; Lee, E.; Kekalo, K.; Liu, J. Design and Optimization of Nanoparticle-Pigmented Solar Selective Absorber Coatings for High-Temperature Concentrating Solar Thermal Systems. *J. Appl. Phys.* **2018**, *123* (3). https://doi.org/10.1063/1.5009252.

(15) Wang, X.; Lee, E.; Xu, C.; Liu, J. High-Efficiency, Air-Stable Manganese–Iron Oxide Nanoparticle-Pigmented Solar Selective Absorber Coatings toward Concentrating Solar Power Systems Operating at 750 °C. *Mater. Today Energy* **2021**, *19*. https://doi.org/10.1016/j.mtener.2020.100609.

(16) Brik, M. G.; Suchocki, A.; Kamińska, A. Lattice Parameters and Stability of the Spinel Compounds in Relation to the Ionic Radii and Electronegativities of Constituting Chemical Elements. *Inorg. Chem.* **2014**, *53* (10), 5088–5099. https://doi.org/10.1021/ic500200a.

(17) Vandenberghe, R. E.; Robbrecht, G. G.; Brabers, V. A. M. On the Stability of the Cubic Spinel Structure in the System Cu-Mn-O. *Mater. Res. Bull.* **1973**, *8* (5), 571–579. https://doi.org/10.1016/0025-5408(73)90134-7.

(18) Liu, Y.; Ying, Y.; Fei, L.; Liu, Y.; Hu, Q.; Zhang, G.; Pang, S. Y.; Lu, W.; Mak, C. L.; Luo, X.; Zhou, L.; Wei, M.; Huang, H. Valence Engineering via Selective Atomic Substitution on Tetrahedral Sites in Spinel Oxide for Highly Enhanced Oxygen Evolution Catalysis. *J. Am. Chem. Soc.* **2020**, *141* (20), 8136–8145. https://doi.org/10.1021/jacs.8b13701.

(19) Navrotsky, A.; Kleppa, O. J. The Thermodynamics of Cation Distributions in Simple Spinels. *J. Inorg. Nucl. Chem.* **1967**, *29* (11), 2701–2714. https://doi.org/10.1016/0022-1902(67)80008-3.

(20) Shoemaker, D. P.; Li, J.; Seshadri, R. Unraveling Atomic Positions in an Oxide Spinel with Two Jahn-Teller Ions: Local Structure Investigation of CuMn2O4. *J. Am. Chem. Soc.* **2009**, *131* (32), 11450–11457. https://doi.org/10.1021/ja902096h.

(21) Lakshmi, S.; Endo, T.; Siva, G. Electronic (Absorption) Spectra of 3d Transition Metal Complexes. *Adv. Asp. Spectrosc.* **2012**, 3–48. https://doi.org/10.5772/48089.

(22) Bosi, F.; Hålenius, U.; Andreozzi, G. B.; Skogby, H.; Lucchesi, S. Structural Refinement and Crystal Chemistry of Mn-Doped Spinel: A Case for Tetrahedrally Coordinated Mn3+ in an Oxygen-Based Structure. *Am. Mineral.* **2007**, *92* (1), 27–33. https://doi.org/10.2138/am.2007.2266.

(23) Huang, R.; Ikuhara, Y. H.; Mizoguchi, T.; Findlay, S. D.; Kuwabara, A.; Fisher, C. A. J.; Moriwake, H.; Oki, H.; Hirayama, T.; Ikuhara, Y. Oxygen-Vacancy Ordering at Surfaces of Lithium Manganese(III,IV) Oxide Spinel Nanoparticles. *Angew. Chemie - Int. Ed.* **2011**, *50* (13), 3053–3057. https://doi.org/10.1002/anie.201004638.

(24) Eppstein, R.; Caspary Toroker, M. On the Interplay Between Oxygen Vacancies and Small Polarons in Manganese Iron Spinel Oxides. *ACS Mater. Au* **2022**. https://doi.org/10.1021/acsmaterialsau.1c00051.

(25) Le Nestour, A.; Gaudon, M.; Villeneuve, G.; Daturi, M.; Andriessen, R.; Demourgues, A. Defects in Divided Zinc-Copper Aluminate Spinels: Structural Features and Optical Absorption Properties. *Inorg. Chem.* **2007**, *46* (10), 4067–4078. https://doi.org/10.1021/ic0624064.

(26) Buvaneswari, G.; Aswathy, V.; Rajakumari, R. Comparison of Color and Optical Absorbance Properties of Divalent Ion Substituted Cu and Zn Aluminate Spinel Oxides Synthesized by Combustion Method towards Pigment Application. *Dye. Pigment.* **2015**, *123*, 413–419. https://doi.org/10.1016/j.dyepig.2015.08.024.

(27) F, R. A.; B, F.; S, H.; H, U.; Aldo, B.; Orabona, E.; Bari, I.-; Università, S.; Moro, P. A.; Roma, I.-. Cation Ordering over Short-Range and Long-Range Scales in the MgAl2O4-CuAl2O4 Series. **2021**, *97* (March), 1821–1827.

(28) Zviagin, V.; Sturm, C.; Esquinazi, P. D.; Grundmann, M.; Schmidt-Grund, R. Control of Magnetic Properties in Spinel ZnFe2O4 Thin Films through Intrinsic Defect Manipulation. *J. Appl. Phys.* **2020**, *128* (16). https://doi.org/10.1063/5.0019712.

(29) Youn, Y.; Miller, J.; Nwe, K.; Hwang, K. J.; Choi, C.; Kim, Y.; Jin, S. Effects of Metal Dopings on CuCr2O4 Pigment for Use in Concentrated Solar Power Solar Selective Coatings. *ACS Appl. Energy Mater.* **2019**, *2* (1), 882–888. https://doi.org/10.1021/acsaem.8b01976.

(30)  Karas, D. E.; Byun, J.; Moon, J.; Jose, C. Copper-Oxide Spinel Absorber Coatings for High-Temperature Concentrated Solar Power Systems. *Sol. Energy Mater. Sol. Cells* **2018**, *182* (March), 321–330. https://doi.org/10.1016/j.solmat.2018.03.025.

(31)  Noč, L.; Ruiz-Zepeda, F.; Merzel, F.; Jerman, I. High-Temperature "Ion Baseball" for Enhancing Concentrated Solar Power Efficiency. *Sol. Energy Mater. Sol. Cells* **2019**, *200* (April). https://doi.org/10.1016/j.solmat.2019.109974.

(32)  Yang, D.; Zhao, X.; Liu, Y.; Li, J.; Liu, H.; Hu, X.; Li, Z.; Zhang, J.; Guo, J.; Chen, Y.; Yang, B. Enhanced Thermal Stability of Solar Selective Absorber Based on Nano-Multilayered AlCrSiO Films. *Sol. Energy Mater. Sol. Cells* **2020**, *207* (December 2019), 110331. https://doi.org/10.1016/j.solmat.2019.110331.

# Supporting Information

Spinel Cu-Mn-Cr Oxide Nanoparticle-Pigmented Solar Selective Coatings Maintaining >94% Efficiency at 750ºC

*Can Xu, Xiaoxin Wang, and Jifeng Liu\**

Thayer School of Engineering, Dartmouth College, 14 Engineering Drive, Hanover, New Hampshire 03755, USA

*Corresponding Author: Jifeng.Liu@dartmouth.edu

## 1. Experimental Methods

**Synthesis.** To obtain Cu, Mn and Cr oxide nanoparticle precursors, copper nitrate ($Cu(NO_3)_2 \cdot 3H_2O$, Fisher Scientific, C99.9), manganese nitrate ($Mn(NO_3)_2 \cdot 4H_2O$, Sigma-Aldrich, C99.5) and chromium nitrate ($Cr(NO_3)_3 \cdot 9H_2O$, Sigma-Aldrich, C99.8), were dissolved in deionized (DI) water at a molar ratio of 1:3:1 at room temperature with an initial pH value between 2 and 3. The aqueous solution was under vigorous magnetic stirring for homogeneity and an excess amount of appropriate base solution, sodium hydroxide (NaOH(aq), 50%, Sigma-Aldrich, dilute to 10M) as we selected, was steadily added dropwise for precipitation until the pH of the solution was adjusted to around 12 for a full precipitation. The stirring was kept for 1h after the precipitation. The precipitated material was rinsed 5 times and then dried at 120 ºC overnight. After a rough grinding step, it was further calcined at 550°C for 5h and finally ground into fine powders.

**Solar Selective Coating Fabrication.** To obtain spraying precursors, 4 wt.% synthesized nanoparticles were well dispersed in xylene diluted silicone resin (BLUESIL RES 6406XM) with a ratio of 1:10 for an appropriate viscosity. The precursors were put in an ultrasonic bath for 30 min and then sprayed onto Inconel 625 substrate on the hot plate with a surface temperature of about 180 ºC. The sample was settled on the hotplate for 5 min and then cooled down to room temperature. Subsequently, the sample was put into a muffle furnace (Thermo scientific) for the following heat treatment. The sample was first heated under 250 ºC for 2h and then ramped up to 750 ºC at a rate of 9.1C/min. After dwelling for 24h and cooling back to room temperature, the nanoparticle-pigmented silicone solar selective coating was formed.

**Thermal test.** A total of 60 simulated day-night thermal cycles (total annealing time of 720h) were conducted for stability tests at 750 ºC and 800 ºC, respectively. Each separate cycle consisted of 10 days. In each day, the sample was heated up to the target temperature at a rate of 9.1C/min and dwelled for 12h. Then the sample was cooled down to room temperature until the start of the next day.

**Characterization.** XRD patterns were recorded on a Rigaku 007 X-ray Diffractometer (Cu K$\alpha$1 line, $\lambda$ = 1.54059 Å) operating at 40 kV/300 mA and in a 2theta angular range of 10–90 degree with a velocity of 2 degree/min and a step size of 0.02 degree. Chemical composition was determined by X-ray Photoelectron Spectroscopy (XPS, PHI VersaProbe II, from West Campus Materials Characterization Core at Yale University) and Energy dispersive spectroscopy (EDS, EDAX Si (Li) detector with Genesis software). Scanning electron images were obtained by using a TESCAN SEM operating at 20 kV in secondary electron (SE) mode and a FEI Helios 5CX DualBeam SEM equipped with FIB was utilized for specimen preparation for cross section views. Tecnai F20 (200 keV) TEM was utilized to collect transmission electron images and SAED patterns. Elemental distribution was measured via JEOL 2010 FEG - TEM/STEM equipped with an EDS detector (from Center for Nanoscale Systems at Harvard University). Vibrational signals of bonding as well as reflectance in the mid infrared (MIR) region (λ=2.5 ~15 μm) were carried out by a Jasco 4100 Fourier transformation IR (FTIR) spectrometer equipped with a Pike IR integrating sphere in the range from 400 to 4000 cm$^{-1}$. Jasco V-570 ultraviolet/visible/near infrared (UV/Vis/NIR) spectrometer equipped with a Jasco ISN-470 integrating sphere was used to characterize optical performance in the ultraviolet, visible and infrared regime, ranging from 200 nm to 2500 nm. The visualization of the surface roughness was obtained via Keyence VHX-700 Digital Microscope.

The solar-to-thermal energy conversion efficiency of the solar selective coatings is given by

$$\eta_{therm} = FOM = \frac{\int(1-R(\lambda))I(\lambda)d\lambda - \frac{1}{C}[\int(1-R(\lambda))B(\lambda,T)d\lambda]}{\int I(\lambda)d\lambda} = \alpha_{solar} - \frac{\varepsilon_{therm}\sigma T^4}{CI_{solar}}, \quad (1)$$

where $R(\lambda)$ is the spectral reflectance of the solar selective coating at wavelength $\lambda$, $I(\lambda)$ represents the AM 1.5 solar spectral irradiance per square meter at wavelength $\lambda$, $I_{solar} = 1000\ W/m^2$ is the solar power density integrated from the spectral radiance $I(\lambda)$, $B(\lambda, T)$ is the spectral blackbody thermal emission at $\lambda$ and $T$, $\alpha_{solar}$ is the overall spectrally normalized solar absorbance, $\varepsilon_{therm}$ is the overall thermal emittance at $T$, and $\sigma$ is the Stefan-Boltzmann constant of $5.67 \times 10^{-8} \frac{W}{m^2 K^4}$, and C=1000 is the solar concentration ratio of power tower CSP systems.

## 2. Nanoparticle Size (Diameter) Histogram

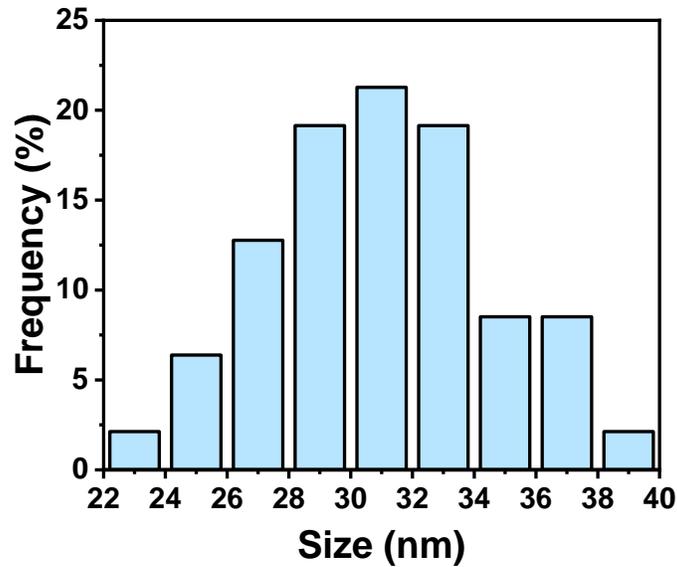

**Figure S1**. Nanoparticle size distribution histogram. The average diameter is 31±3.6 nm.

## 3. X-ray Photoelectron Spectroscopy (XPS) Data Analyses

Figure S2.a shows a survey spectrum of as-synthesized Cu-Mn-Cr oxide nanoparticles. Cu, Mn, Cr and O are detected from the surface. Figure S2.b shows the binding energies of the Cu 2p core levels. Two sharp peaks at about 932 eV and 952 eV are observed, corresponding to the significantly split spin-orbit components Cu $2p_{3/2}$ and Cu $2p_{1/2}$ respectively. Shake-up structures at about 945 eV and 963 eV are satellite features of Cu with oxidation states. Such two satellite peaks reveal the existence of Cu(II) while the relatively low intensity indicates possible mixed states of Cu(I) and Cu(II).

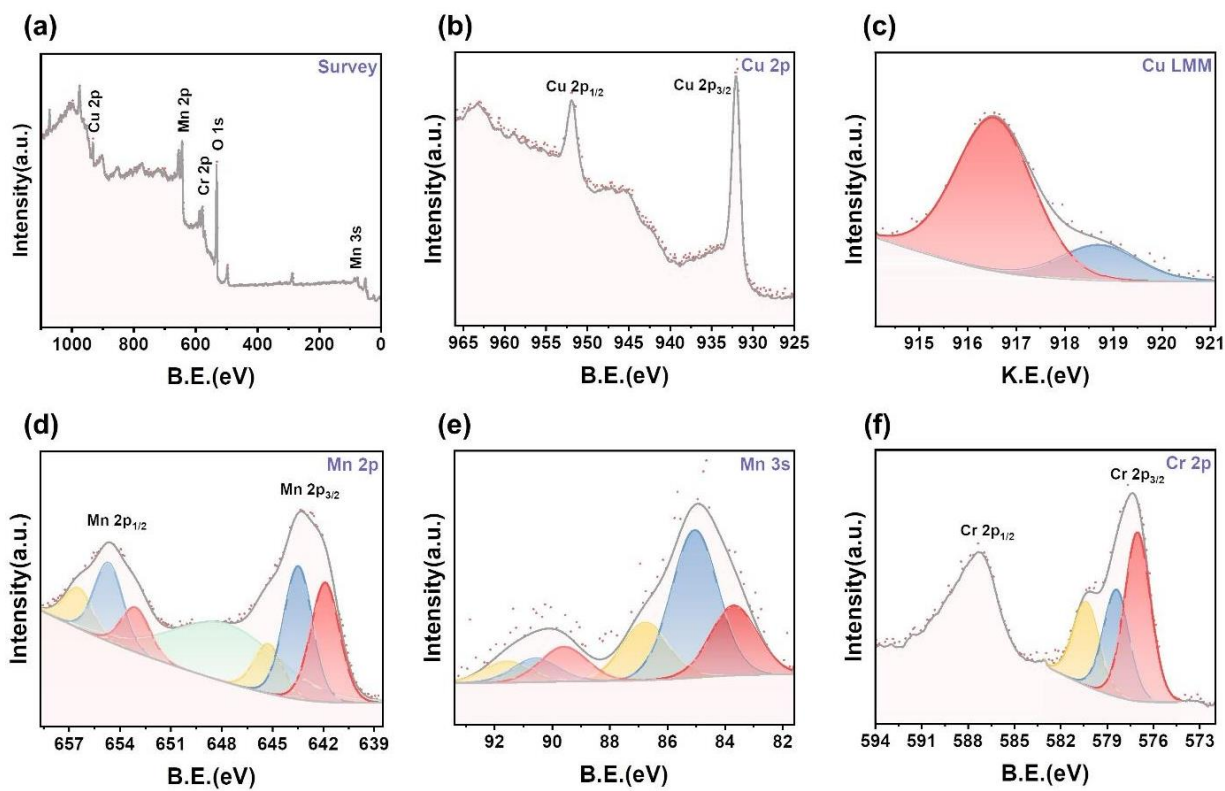

**Figure S2**. XPS spectra of Cu, Mn, and Cr ions in the synthesized spinel oxide nanoparticles.

To further investigate the oxidation states and their compositions, principal Cu LMM Auger peaks (Figure S2.c) are collected as well. One deconvoluted peak with a binding energy of 916.5 eV is assigned to Cu(I) [1] and the other one at 918.4 eV is assigned to Cu(II) [2] with an error of ±1 eV. The modified Auger parameters are calculated and compared to minimize the effect of charging of non-conducting specimens during the measurement. Cu in $CuCr_2O_4$ has an Auger parameter of nearly 1853 eV [3], close to 1852.7 eV, the value we obtained for Cu(II). Biesinger [4] analyzed XPS data of various copper-containing species in previously published literature and summarized their Auger parameters. Our Cu(I) has an Auger parameter of 1848.6 eV, which lies in the range of several Cu(I) involved materials. In this case, Cu(I) and Cu(II) are present simultaneously with a ratio of about 4 : 1, taking into account both the percentage of peak areas and Relative Sensitivity Factors (RSF).

Mn 2p (Figure S2.d) and Mn 3s (Figure S2.e) data are collected for quantification and qualification of the chemical species in the specimen. In the 2p spectrum, two peaks at about 643.0 eV and 654.5 eV are assigned to the Mn $2p_{3/2}$ and Mn $2p_{1/2}$ components. An extremely broad and weak satellite peak observed at around 649.1 eV is the feature of Mn(II). Three deconvolved peaks represent Mn(II), Mn(III) and Mn(IV) with increasing values of binding energies. Mn 3s spectrum distinguishes Mn oxidation states in a more straightforward way. Based on the photoemission final states with the s electrons parallel or antiparallel to the 3d spin [5], the fewer and fewer 3d unpaired electrons in Mn(II), Mn(III) and Mn(IV) lead to the smaller and smaller magnitudes of peak splitting, from around 6.0 eV to 5.4 eV and to 4.7 eV for oxides [6]. This assists in identifying the oxidation states and calculating the percentage respectively by taking the energy difference as a new constraint during the curve fitting. Eventually, Mn(II), Mn(III) and Mn(IV) are confirmed coexisting in our specimen with a percentage of 30%, 47% and 23%. Figure 2.f shows the Cr 2p spectrum, with Cr $2p_{3/2}$ peak at around 577.5 eV and Cr $2p_{1/2}$ peak at around 587.3 eV. Cr in $CuCr_2O_4$ is located at 577.3 eV [2] and that suggests the existence of Cr(III) [7,8]. Three sub-peaks with the same full width at half maximum (FWHM) around 1.979eV are deconvoluted for Cr $2p_{3/2}$ peak, showing the multiplet structure [9] due to the coupling effect between the unpaired core electron and unpaired electrons in the outer shell [10] in Cr(III).

## 4. Fourier Transform Infrared Spectroscopy (FTIR)

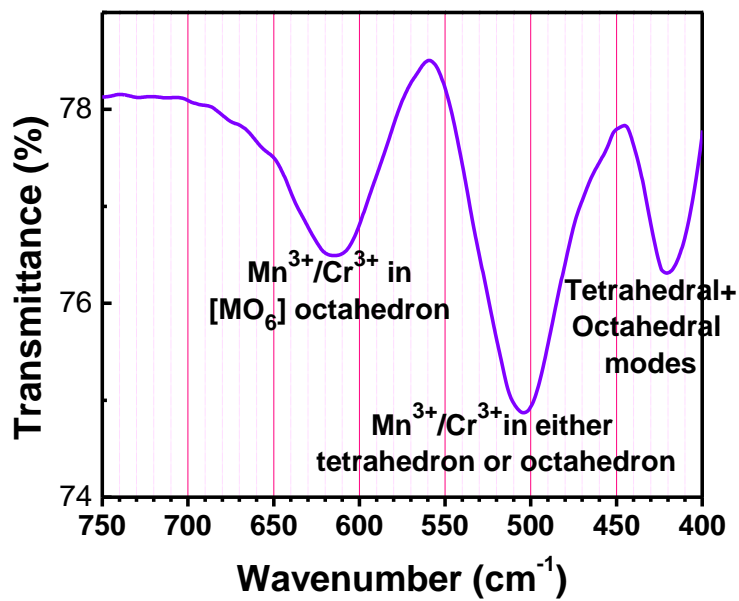

**Figure S3**. FTIR spectrum of the Cu-Mn-Cr oxide nanoparticles showing characteristic spinel features.

Characteristic IR absorption peaks are observed at ~616, 505, and 420 cm$^{-1}$, close to the previous reports on spinel $CuMn_2O_4$ [11], $ZnMn_2O_4$ [12], and $MnCr_2O_4$ [13]. According to Ref. 13, the peak at ~616 cm$^{-1}$ corresponds to trivalent cation vibrations in the [$MO_6$] octahedron. In our case this peak is asymmetric, which can be induced by the coexistence of $Mn^{3+}$ and $Cr^{3+}$ as well as valance 2+ and 4+ cations on the octahedral sites, as revealed by the XRD, EDS and XPS analyses in the main text. The second peak at 505 cm$^{-1}$ is attributed to trivalent ions, either on tetrahedral or octahedral sites. The last peak at 420 cm$^{-1}$ is a complex vibrational mode involving both tetrahedral and octahedral sites.

## 5. Nanoparticle Volume Fraction Estimation

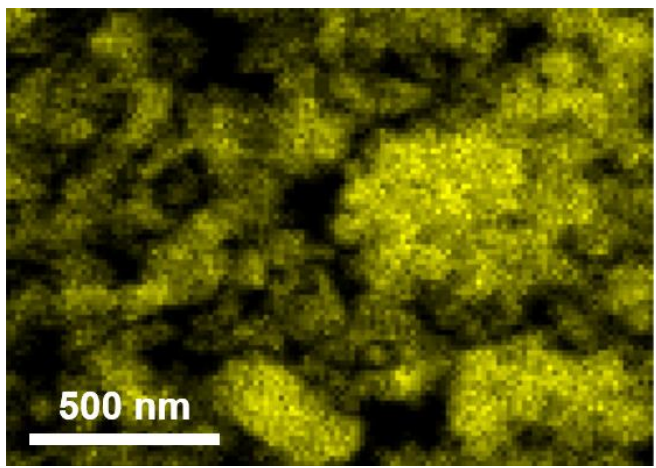

**Figure S4**. EDS mapping of Mn in a cross-sectional FIB lamellar of the coating. The lamellar is ~150 nm thick.

Since Mn is only present in the pigment nanoparticles and not in the silicone matrix, the Inconel substrate, or the TEM grid (which is used to mount the FIB sample of the coating), we can use Mn as a characteristic element of the NPs to derive the corresponding volume fraction by analyzing the EDS mapping of Mn (Figure S4). We utilized Image J to obtain the area fraction of Mn in the selected area. As shown in Figure S4, the yellow dots represent Mn signals and a brighter color reveals a higher Mn intensity in that region. The original figure was firstly converted to an 8-bit binary image and a threshold ranging from 51 to 255 was chosen to select the Mn pixels. Mn is determined to take 48.0% of the entire selected area.

Assuming spherical nanoparticle approximation and no overlapping of the nanoparticles in the vertical direction, the volume fraction could be expressed as the following equation:

$$f = \frac{V_{Mn}}{V_{total}} = \frac{\frac{A*x}{\pi r^2} \times \frac{4}{3}\pi r^3}{A \times t} = \frac{4}{3}\left(\frac{r}{t}\right) * x, \qquad (2)$$

where $x$ is the area fraction, $r$=15.5±1.8 nm is the radius of nanoparticles (see the histogram in Figure S1) and $t = 150\ nm$ is the thickness of the FIB processed lamellar. With this equation, a volume fraction of $f$ ~7 vol. % was derived, which could be regarded as the lower limit since nanoparticles do overlap in reality.

Another estimation was conducted according to the parameters used during the fabrication process, including the weight percentage of nanoparticles in the precursor, the density of each nonvolatile component, and the average coating thickness of 8.5 μm (as discussed in the main text). Assuming no excessive loss of nanoparticles during the coating process and annealing process, we obtain an upper limit of volume fraction $f$ ~13 vol. %. Taking the average between the lower limit estimated by EDS Mn area mapping and the upper limit from chemical precursor ratios, it is reasonable to consider the volume fraction $f$=10±3 vol. % for comparison with theoretical modeling.

## 6. XRD Analyses during Thermal Endurance Tests

XRD measurements were taken every 10 day-night annealing cycles, and Figures S5a and 5c were plotted with each intermediate state during the whole thermal cycle procedure at 750 ℃ and 800 ℃ separately to show the deviation for several important peaks. Generally, the cycled coating shows similar X-ray diffraction patterns while minor shifts occur. The peak position of the characteristic spinel (311) peak is closely examined as shown in Figure S5.b and 5d for 750 ℃ and 800 ℃, respectively. It originally lies at 35.78° and tends to shift as the thermal cycle starts. It stabilizes at 35.62° for 750 ℃ and 35.56° for 800 ℃ after 20 day-night simulated cycles. Based on Bragg's equation, the peak shift towards a smaller diffraction angle refers to a larger interplanar spacing, which means the lattice expands slightly.

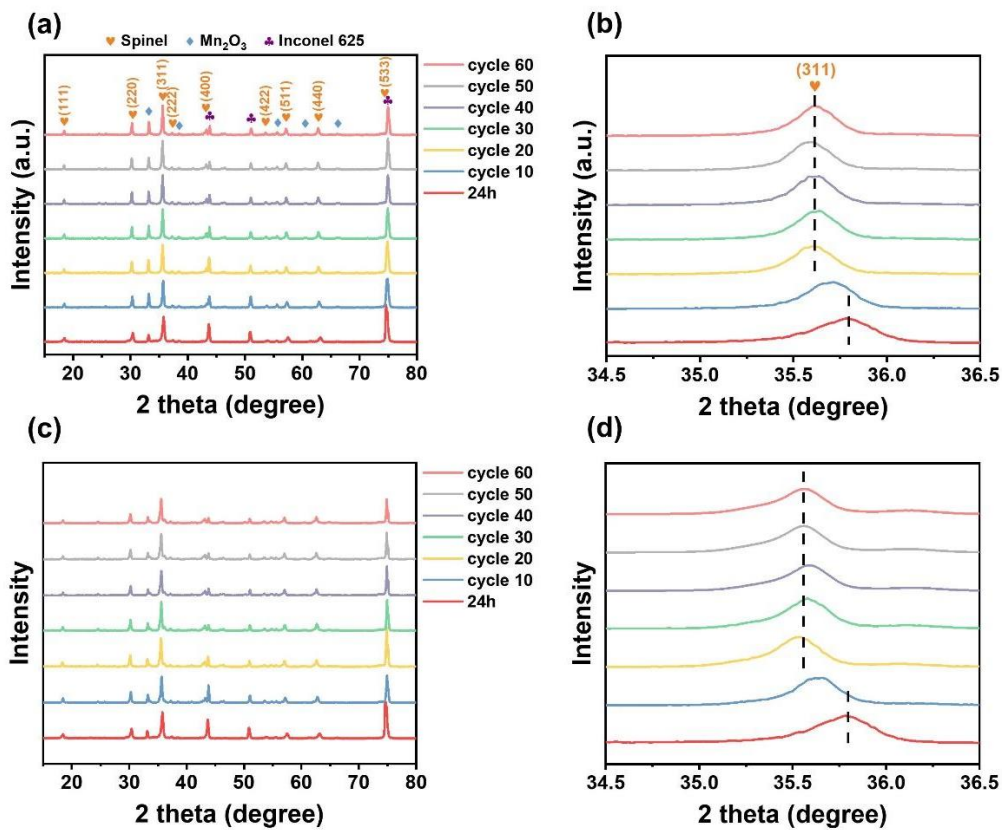

**Figure S5**. XRD patterns during thermal endurance tests at (a) 750 °C and (b) 800 °C, with close position examination of spinel (311) peak for samples through thermal cycles at (c) 750 °C and (d) 800 °C, respectively.

From previous work published by Mikhail G. Brik [14], the lattice constants of $CuMn_2O_4$ and $CuCr_2O_4$ are 8.33 Å and 8.27 Å, while that of $MnCr_2O_4$ is 8.437 Å. A variation from 8.410 Å to 8.474 Å depending on different milling hours was reported by R. N. Bhowmik[15]. Considering the valences and atom position distributions discussed in the previous section, as more Cr ions dope into the spinel system and take the octahedral site, Cu and Mn ions tend to sit in the tetrahedral site, making it closer to the structure of $CuCr_2O_4$ and $MnCr_2O_4$. Taking into account the lattice parameters mentioned above and ion radius data obtained from WebElements [16], it is reasonable to observe the lattice expansion. Mn ions are partially released to form $Mn_2O_3$, which is in agreement with a higher $Mn_2O_3$/spinel ratio (from 0.142:1 to 0.466:1, stabilizing after 30 day-night thermal cycles at 750 ºC) derived from XRD result.

## 7. Optical Spectra Evolution and EDS Mapping for Interdiffusion Investigation upon Thermal Cycling

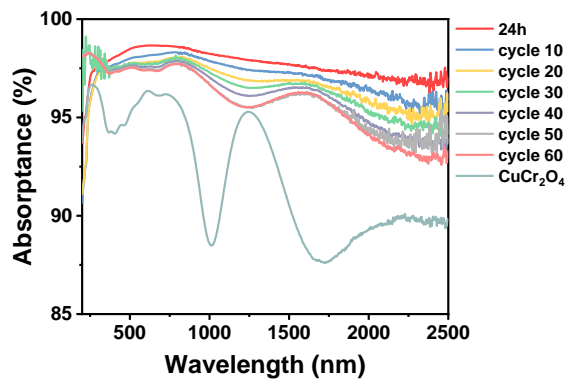

**Figure S6**. UV-vis-NIR absorption spectra of Cu-Mn-Cr oxide nanoparticle pigmented solar selective coating during thermal endurance tests at 800 ºC compared with as-coated $CuCr_2O_4$ nanoparticle pigmented coating.

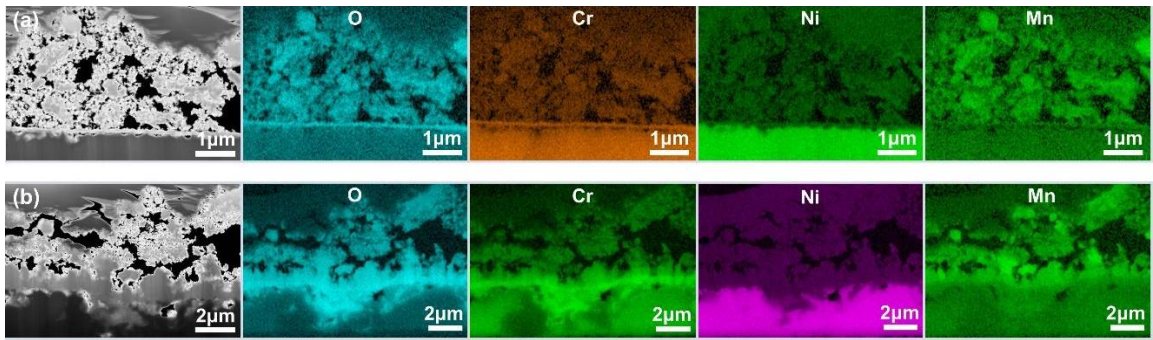

**Figure S7**. Cross-section STEM image and EDS mapping result of cross-sectional FIB cut specimens of Cu-Mn-Cr oxide nanoparticle pigmented solar selective coating (a) before and (b)

after 60 day-night thermal cycles at 750 ºC. Part of the coating was damaged during the FIB milling processed.

Detailed investigation in the interface between the nanoparticle-pigmented coating and the Inconel alloy substrate was conducted by observing the cross-sections of the specimen processed by FIB milling. According to the STEM image shown in Figure S7a, an oxide layer of around 100 nm was formed after 24h annealing at 750 ºC. Further EDS mapping reveals that the oxide layer mainly consists of Cr based oxides. As an obvious comparison in Figure S7b, a much thicker oxide layer was observed for the sample that has completed 60 day-night simulated cycles at 750 ºC. It approximately increases to 1 μm thick after long-time thermal cycles. This is a common oxide scale when oxidizing Inconel alloys.

EDS analyses of the coating were also conducted, and an increasing Cr concentration with thermal cycling was detected. Near the surface of the coatings, the Cr concentration increased by 58% after 60 day-night cycles at 750 ºC, and 117% after 60 day-night cycles at 800 ºC. This clearly demonstrates that Cr atoms have diffused from the Inconel substrate through the solar coating and emerged at the upper layers of the coating. The observed Cr diffusion into the coating is fully consistent with the optical absorption spectrum evolution shown in Figure S6 and the corresponding discussions in the main text.

## 8. Spray-Coated Solar Selective Coating on 48-Inch-Long Tube

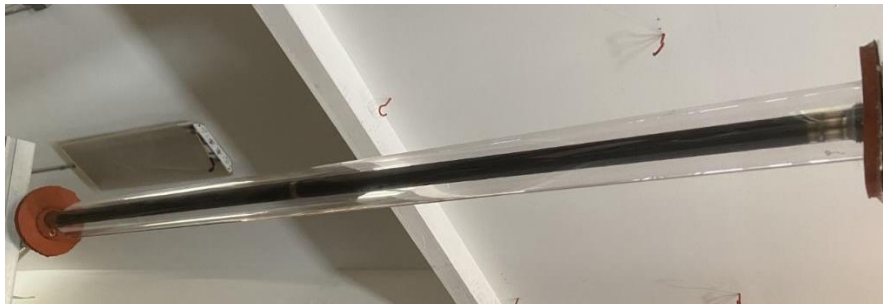

**Figure S8**. A photo of spinel Cu-Mn-Cr oxide nanoparticle pigmented solar selective coating on a 48-inch-long tube prepared by spray coating method.

## Reference

(1) Losev, A.; Rostov, K.; Tyuliev, G. Electron Beam Induced Reduction of CuO in the Presence of a Surface Carbonaceous Layer: An XPS/HREELS Study. *Surf. Sci.* **1989**, *213* (2–3), 564–579. https://doi.org/10.1016/0039-6028(89)90313-0.

(2) Capece, F. M.; Castro, V. Di; Furlani, C.; Mattogno, G.; Fragale, C.; Gargano, M.; Rossi, M. "Copper Chromite" Catalysts: XPS Structure Elucidation and Correlation with Catalytic Activity. *J. Electron Spectros. Relat. Phenomena* **1982**, *27* (2), 119–128. https://doi.org/10.1016/0368-2048(82)85058-5.

(3) NIST X-ray Photoelectron Spectroscopy Database, NIST Standard Reference Database Number 20, National Institute of Standards and Technology, Gaithersburg MD, 20899 (2000) http://dx.doi.org/10.18434/T4T88K (accessed 2021 -10 -13).

(4)     Biesinger, M. C. Advanced Analysis of Copper X-Ray Photoelectron Spectra. *Surf. Interface Anal.* **2017**, *49* (13), 1325–1334. https://doi.org/10.1002/sia.6239.

(5)     Waskowska, A.; Gerward, L.; Olsen, J. S.; Steenstrup, S.; Talik, E. CuMn2O4: Properties and the High-Pressure Induced Jahn-Teller Phase Transition. *J. Phys. Condens. Matter* **2001**, *13* (11), 2549–2562. https://doi.org/10.1088/0953-8984/13/11/311.

(6)     Junta, J. L.; Hochella, M. F. Manganese (II) Oxidation at Mineral Surfaces: A Microscopic and Spectroscopic Study. *Geochim. Cosmochim. Acta* **1994**, *58* (22), 4985–4999. https://doi.org/10.1016/0016-7037(94)90226-7.

(7)     Biesinger, M. C.; Payne, B. P.; Grosvenor, A. P.; Lau, L. W. M.; Gerson, A. R.; Smart, R. S. C. Resolving Surface Chemical States in XPS Analysis of First Row Transition Metals, Oxides and Hydroxides: Cr, Mn, Fe, Co and Ni. *Appl. Surf. Sci.* **2011**, *257* (7), 2717–2730. https://doi.org/10.1016/j.apsusc.2010.10.051.

(8)     Mohamed, R. M.; Kadi, M. W. Generation of Hydrogen Gas Using CuCr2O4-g-C3N4 Nanocomposites under Illumination by Visible Light. *ACS Omega* **2021**, *6* (6), 4485–4494. https://doi.org/10.1021/acsomega.0c06193.

(9)     Oladipo, A. A. Rapid Photocatalytic Treatment of High-Strength Olive Mill Wastewater by Sunlight and UV-Induced CuCr2O4@CaFe–LDO. *J. Water Process Eng.* **2021**, *40* (January), 2–6. https://doi.org/10.1016/j.jwpe.2021.101932.

(10)    Moulder, J. F.; Stickle, W. F.; E.'Sobol, P.; Bomben, K. D. *Handbook of X-Ray Photoelectron Spectroscopy*; Perkin-Elmer Corp, Eden Prairie, MN, 1992. https://doi.org/10.1002/0470014229.ch22.

(11) Patra, P.; Naik, I.; Kaushik, S. D.; Mohanta, S. Crossover from Meta-Magnetic State to Spin-Glass Behaviour upon Ti-Substitution for Mn in CuMn2O4. *J. Mater. Sci. Mater. Electron.* **2021**, *33* (1), 554–564. https://doi.org/10.1063/5.0017104.

(12) Ghannam, M. M.; Heiba, Z. K.; Sanad, M. M. S.; Mohamed, M. B. Functional Properties of ZnMn2O4/MWCNT/Graphene Nanocomposite as Anode Material for Li-Ion Batteries. *Appl. Phys. A Mater. Sci. Process.* **2020**, *126* (5), 1–9. https://doi.org/10.1007/s00339-020-03513-6.

(13) Allen, G. C.; Paul, M. Chemical Characterization of Transition Metal Spinel-Type Oxides by Infrared Spectroscopy. *Appl. Spectrosc.* **1995**, *49* (4), 451–458. https://doi.org/10.1366/0003702953964372.

(14) Brik, M. G.; Suchocki, A.; Kamińska, A. Lattice Parameters and Stability of the Spinel Compounds in Relation to the Ionic Radii and Electronegativities of Constituting Chemical Elements. *Inorg. Chem.* **2014**, *53* (10), 5088–5099. https://doi.org/10.1021/ic500200a.

(15) Bhowmik, R. N.; Ranganathan, R.; Nagarajan, R. Lattice Expansion and Noncollinear to Collinear Ferrimagnetic Order in a MnCr2O4 Nanoparticle. *Phys. Rev. B - Condens. Matter Mater. Phys.* **2006**, *73* (14), 1–9. https://doi.org/10.1103/PhysRevB.73.144413.

(16) WebElements https://www.webelements.com (accessed 2022 -01 -22).